\documentclass[twocolumn,preprintnumbers,amsmath,amssymb,superscriptaddress]{revtex4}
\usepackage{graphicx}
\usepackage{dcolumn}
\usepackage{bm}
\usepackage{soul}
\usepackage{color}
\usepackage{epstopdf}
\usepackage[version=3]{mhchem}
\usepackage{lipsum}
\usepackage[outercaption]{sidecap}
\usepackage{floatrow}
\usepackage{hyperref}

\begin{document}

%\preprint{APS/123-QED}

\title{Enhanced solar photothermal effect of PANi fabrics with plasmonic nanostructures}
\author{Do T. Nga}
\affiliation{Institute of Physics, Vietnam Academy of Science and Technology, 10 Dao Tan, Ba Dinh, Hanoi 10000, Vietnam}
\email{dtnga@iop.vast.ac.vn}
\author{Anh D. Phan}
\email{anh.phanduc@phenikaa-uni.edu.vn}
\affiliation{Phenikaa Institute for Advanced Study, Artificial Intelligence Laboratory, Faculty of Computer Science, Materials Science and Engineering, Phenikaa University, Hanoi 12116, Vietnam}
\affiliation{Department of Nanotechnology for Sustainable Energy, School of Science and Technology, Kwansei Gakuin University, Sanda, Hyogo 669-1337, Japan}
\author{Vu D. Lam}
\affiliation{Graduate University of Science and Technology, Vietnam Academy of Science and Technology, 18 Hoang Quoc Viet, Hanoi, Vietnam}
\author{Lilia M. Woods}
\affiliation{Department of Physics, University of South Florida, Tampa, Florida 33620, United States}
\author{Katsunori Wakabayashi}
\affiliation{Department of Nanotechnology for Sustainable Energy, School of Science and Technology, Kwansei Gakuin University, Sanda, Hyogo 669-1337, Japan}
\date{\today}

\begin{abstract}
The photothermal energy conversion in hanging and floating polyaniline (PANi)-cotton fabrics is investigated using a model based on the heat diffusion equation. Perfect absorption and anti-reflection of wet hanging PANi-cotton fabrics cause quick transfer of total incident light into water confining nearly 100 $\%$ of the sunlight. As a result, a hanging membrane is found to have more attractive properties than a floating above water fabric. We find, however, that the photothermal properties of a floating PANi-cotton membrane can greatly be enhanced by dispersing TiN nanoparticles in the water below the fabric. The calculated temperature gradients for TiN nanoparticle solutions show that the absorbed energy grows with increasing the nanoparticle density and that  the photothermal process occurs mostly near the surface. The collective heating effects depend on the size and density of nanoparticles, which can further be used to modulate the photothermal process. 
%Due to the densification of electromagnetic fields and two surfaces of vaporization, the hanging fabric has higher surface temperature and vaporizes more water than the floating system without TiN nanoparticles.
\end{abstract}

%Use showkeys class option if keyword
                              %display desired
\maketitle
\section{Introduction}
Photothermal effects have been studied intensively because of their wide range of applications for efficient solar vapor generation \cite{17}, optical data storage \cite{18}, alternative cancer therapy \cite{21,22}, antibacterial activities \cite{23}, and radiative cooling \cite{19}. Under illumination of light, free electrons are collectively excited on the surface of photothermal agents. Due to surface plasmon resonance, which provides the abilities of structures to confine and enhance electromagnetic fields, the absorbed optical energy thermally dissipates into the surroundings and increases the temperature of the medium. To tailor photothermal phenomena for desired purposes, one often takes advantage of the spatial inhomogeneity of the electric fields and temperature distributions in the electron density. Additionally, it has been shown that noble metals and alternative plasmonic materials provide high efficiency of light-to-heat conversion \cite{24,25}. Also recently, researchers have started exploring environmentally friendly polymers, different types of wood, and non-metal materials in photothermal applications \cite{6}.

Finding ways to use solar energy effectively is at the research forefront currently since renewable sources are needed for energy consumption. However, enhancing solar energy storage capabilities and increasing light-to-heat conversion efficiency are challenging. To avoid heat losses in a photothermal process, one of the most promising strategies is a direct energy transfer from the Sun to water for seawater desalination. Apart from using broadband solar perfect absorbers \cite{6}, it is possible to use nanostructures \cite{31,26,16,7,17,15,32} to improve solar energy absorption and increase water heating and steam generation. Several solar photothermal systems \cite{31,26,16,7,17,15,32} consist of dispersed nanostructures, and submerged and floating plasmonic membranes, which are made of porous materials, wood, plasmonic nanoparticles, and graphene oxide \cite{32,33}.

The theoretical understanding of solar steam generation and energy conversion has been advanced by investigating different scenarios. For example, Neumann {\it et. al} have used a simple form of the heat diffusion equation to describe phenomenologically the photothermal formation of air bubbles around gold nanoshells in a solution \cite{28}. Other researchers have used finite element analysis to solve heat transfer equations to determine temperature distributions of a single plasmonic structure on glass substrate under simulated solar irradiation \cite{1}. In recent works \cite{31, 15}, we have investigated new plasmonic heating models to calculate time-dependent temperature gradients of gold nanoshell and titanium nitride (TiN) nanoparticle solutions under solar irradiation. TiN has been known as an alternative plasmonic material, which overcomes several limitations of noble-metal materials (like gold) in terms of hardness, chemical and physical stability, a high melting point, and biocompatibility \cite{15}. Our models capture the collective effects of the nanoparticles and can quantitatively describe, for the first time, the vaporization processes for a wide range of particle concentrations. Although our theoretical results agree well with experiments \cite{31,15}, their validity is limited by the fact that the solutions were considered to be uniform with a uniform distribution of nanoparticles. 

In this work, the solar heating of hanging and floating on the surface of water polyaniline-cotton fabrics is theoretically investigated. Polyaniline (PANi) is an environmentally friendly low cost conducting polymer. The fabrics can be used in electronic packaging areas \cite{34,35,36}. In photothermal applications, some of its attractive properties are good conductivity, high absorption in the near infrared regime, and lack of toxicity. Recently, it was reported that a hanging fabric has capabilities for denser confinement of sunlight energy as compared to a floating one on seawater \cite{6}. The higher densification of energy leads to a greater temperature rise and enhanced photothermal conversion. This has been explained by the less thermal dissipation into the medium underneath the floating cotton and presence of two free surfaces of the hanging one leading to acceleration of evaporation. Our theoretical model provides qualitative and quantitative description of the results in Ref. \cite{6} and it shows how the photothermal capabilities of the floating fabric can be much increased using plasmonic nanoparticles dispersed in the water. These nanoparticles absorb sunlight energy and effectively convert this to heat. By calculating steady-state thermal gradients in these two systems including TiN nanoparticles randomly dispersed into solutions below the floating fabric, we find that the enhancement of solar energy harvesting depends on the density of the nanoparticles and their size. Remarkably, this approach can be applied to investigate photothermal responses of multilayered systems.

%%%%%%%%%%%%%%%%%%%%%%%%%%%%%%
\section{Theoretical background}
\begin{figure}[htp]
\includegraphics[width=8.5cm]{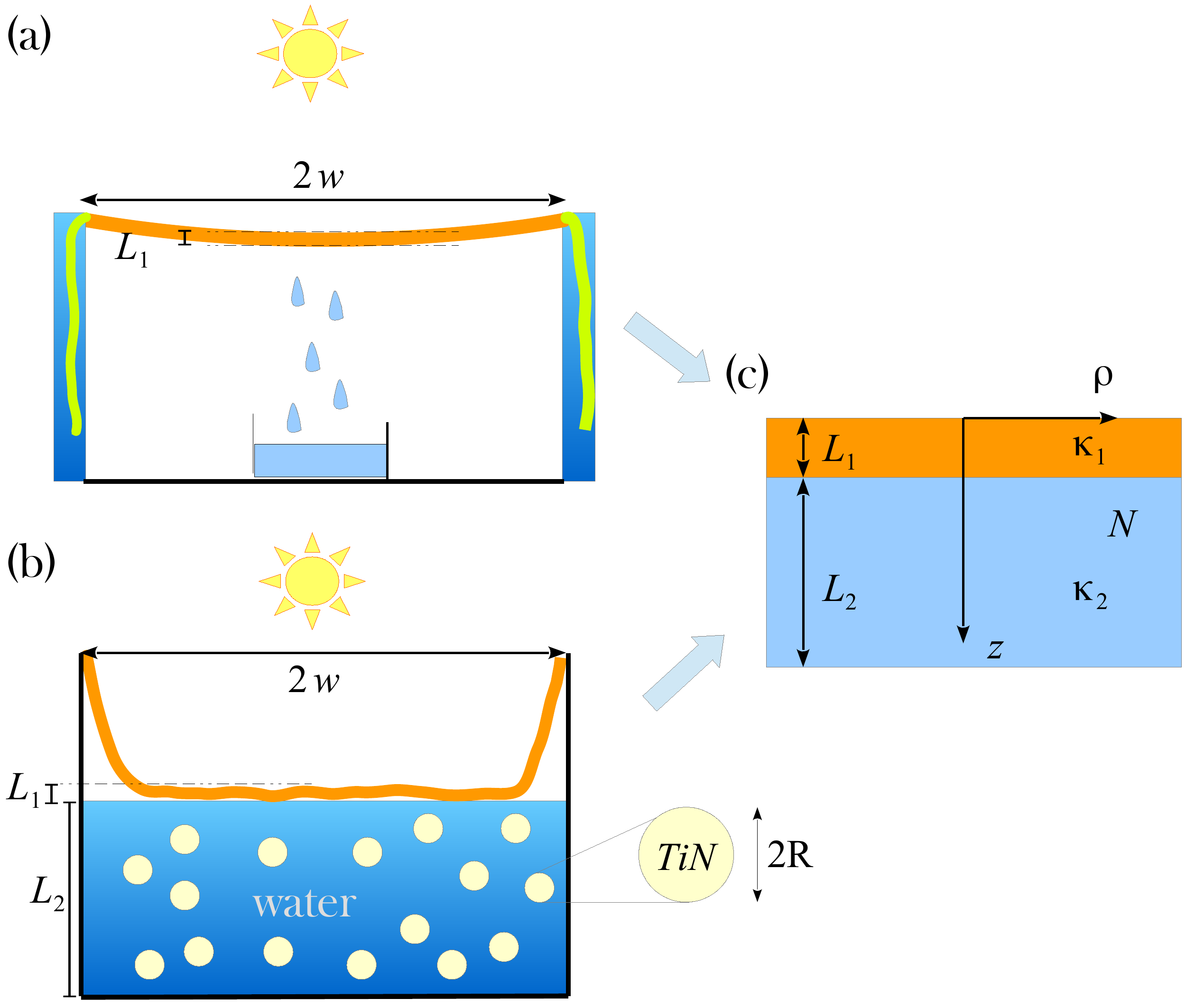}
\caption{\label{fig:1}(Color online) Schematic illustration of (a) the hanging PANi fabric membrane; (b) floating PANi fabric membrane; (c) a simplified layer-like representation, where the thickness $L_1, L_2$ and thermal conductivity $\kappa_1, \kappa_2$ of each layer are shown. The corresponding layer thicknesses for the considered structures are shown for the  hanging and floating membranes. We take $\kappa_1=0.637$ $Wm^{-1}K^{-1}$ and $\kappa_2=0.6$ $Wm^{-1}K^{-1}$, $L_1= 2$ mm, $L_2=0$ for panel (a), and $L_2=3$ cm for panel (b), as reported in \cite{6}. The TiN nanoparticles in panel (b) are randomly dispersed and have spherical shape with radius $R$. The simulated sunlight is uniform and has a spot size of 8 cm $\times$ 8 cm \cite{6}. The effective projected diameter of the floating and hanging fabrics is $2w \approx$ 6.324 cm, which is less than the light spot size of 8 cm. $N$ is the density number of particles, and $\rho, z$ are the cylindrical coordinates.}
\end{figure}

A schematic representation of the systems under consideration is shown in Fig. \ref{fig:1}. A wet by seawater hanging PANi-cotton fabric (Fig. \ref{fig:1}a) is attached by the edges of two vertical walls. The hanging cotton is exposed to the Sun which increases its temperature and leads to vaporization of the  seawater. As a result, the water steam then condenses to form clean water droplets falling down into a container under it. The floating PANi-cotton membrane (Fig. \ref{fig:1}b) rests on top of a layer of seawater which contains randomly dispersed TiN nanoparticles. The fabric enables passing of sunlight through it without reflection heating up the entire seawater TiN nanoparticle solution. The water steam generation occurs at the liquid-air interface instead of up and down surfaces of the hanging system. Although the physical mechanisms of solar driven photothermal effects in these two systems are different, one can give a common two layer structural representation (Fig. \ref{fig:1}c). Specifically, the layer with thickness $L_1$ denotes the layer containing the fabric cotton, while the layer with thickness $L_2$ denotes the solution layer underneath the fabric. For clarity, $L_1$ and $L_2$ are also shown in Figs. \ref{fig:1}a and \ref{fig:1}b. The corresponding for each layer thermal conductivities $\kappa_1$ and $\kappa_2$ are also shown. The sunlight, assumed to be uniform, heats up a spot whose size is  8 cm $\times$ 8 cm, as reported in \cite{6}.

For the system in Fig. \ref{fig:1}b, the effective thermal conductivity of the $L_2$ layer, which contains the aqueous TiN particle solution, is calculated by
\begin{eqnarray}
\kappa_{eff}=\kappa_2(1-\eta) + \kappa_{TiN}\eta,
\label{eq:4-1}
\end{eqnarray}
where $\kappa_{TiN} \approx 29$ $Wm^{-1}K^{-1}$ is the thermal conductivity of bulk TiN \cite{9} and $\eta = 4\pi N R^3/3$ is the volume fraction of TiN nanoparticles. Here, we approximate $\kappa_{eff}$ in a mean-field manner. For the particle density $N$ ranging from $10^{10}$ to $10^{18}$ \ce{nanoparticles/m^3}, $\kappa_{eff} \approx \kappa_2$.

We note that the hanging PANi-cotton membrane has stronger photothermal effects than the floating one. This is due to a vaporization process facilitated by the fact that the hanging cloth has air under and above it. To further enhance the photothermal process in the hanging PANi fabric is take advantage of plasmonic properties of metallic nanoparticles \cite{26,16,7,17,15}. However, incorporating metal-like materials inside fabric reduces its anti-reflective properties, which leads to decreasing the temperature rise and worsening of the conversion process. However, the solar thermal heating process of the floating fabric system can be improved by using plasmonic nanoparticles dispersed in the water.  While most such nanoparticles are made of noble metals, recently TiN systems have been shown as attractive alternatives due to their fabrication and stability as well as greater tunability \cite{31,11,27}. Here we consider TiN spherical nanoparticles randomly dispersed in the water under the floating PANi-cotton membrane (Fig. \ref{fig:1}b) in order to explore possibilities of enhancing its photothermal process.

The temperature rise is governed by the heat diffusion equation, which in cylindrical coordinates ($\rho, z$) is written as, 
\begin{eqnarray}
\kappa_{1,2}\left[\frac{1}{\rho}\frac{d}{d\rho}\left(\rho \frac{d \Delta T}{d\rho}\right) + \frac{d^2\Delta T}{dz^2}\right]=p(\rho,z),
\label{eq:5}
\end{eqnarray}
where $\Delta T \equiv \Delta T(\rho,z)$ is the temperature change and $p(\rho,z)$ is the power or energy absorbed per unit volume. Since the PANi-cotton fabric is a perfect absorber \cite{6}, the absorption of the solar incident flux occurs at the surface ($z=0$) of the substance and there is no heat source meaning that $p(\rho,z)=0$ in each layer. Thus Eq.(\ref{eq:5}) becomes
\begin{eqnarray}
\frac{1}{\rho}\frac{d}{d\rho}\left(\rho \frac{d \Delta T}{d\rho}\right) + \frac{d^2\Delta T}{dz^2}=0.
\label{eq:6}
\end{eqnarray}

By using Hankel transformation of $\Delta T(\rho,z)$ ($\Delta T(\rho,z)=\int_0^\infty\Theta(u,z)J_0(\rho u)udu$ where $J_0(\rho u)$ is the Bessel function of the first kind) \cite{8}  Eq.(\ref{eq:6}) is re-written as
\begin{eqnarray}
-u^2\Theta(u,z) +\frac{d^2\Theta(u,z)}{dz^2} = 0.
\label{eq:7}
\end{eqnarray}
and its solution is 
\begin{eqnarray}
\Theta_{1,2}(u,z)=A_{1,2}(u)e^{-uz}+B_{1,2}(u)e^{uz},
\label{eq:8}
\end{eqnarray}
where $A_1(u)$, $B_1(u)$, $A_2(u)$, and $B_2(u)$ are parameters determined by boundary conditions, which include continuity of the heat flux and temperature at surfaces. For the system in Fig. \ref{fig:1}c, one finds
\begin{eqnarray}
-\kappa_{1}\left.\frac{\partial \Theta(u,z)}{\partial z}\right|_{z=0} &=& \Psi(u), \nonumber\\
\Theta(u,L_1^-) &=& \Theta(u,L_1^+), \nonumber\\
-\kappa_{1}\left.\frac{\partial \Theta(u,z)}{\partial z}\right|_{z=L^-} &=& -\kappa_{2}\left.\frac{\partial \Theta(u,z)}{\partial z}\right|_{z=L_1^+}, \nonumber\\
-\kappa_{2}\left.\frac{\partial \Theta(u,z)}{\partial z}\right|_{z=(L_1+L_2)} &=& h\Theta(u,z=L_1+L_2),
\label{eq:11}
\end{eqnarray}
where $h$ is the convective heat transfer coefficient and $\Psi(u)$ is the Hankel transform of the incoming light flux. Since the spot size of sunlight is greater than the effective size of fabrics $w$, the absorbed solar light intensity is taken to be $P_0\neq 0$ for $\rho \leq w$ and 0 for $\rho \geq w$. The Hankel transform of the incident intensity is
\begin{eqnarray}
\Psi(u) = \int_0^{w} P_0J_0(\rho u)\rho d\rho = P_0\frac{wJ_1(wu)}{u}.
\label{eq:9}
\end{eqnarray}

By solving numerically Eqs. (\ref{eq:11}) and (\ref{eq:9}) and taking the inverse Hankel transform, one obtains $\Delta T(\rho,z)$. According to \cite{6}, the hanging photothermal fabric absorbs  100 $\%$ of light at all wavelengths. Further considering that the authors in Ref. \cite{6} have used a standard solar simulator (Oriel Newport 69911) with 300 to 2500 nm wavelength range, we find that $P_ 0 \approx \int_{300}^{2500}E_\lambda d\lambda \approx 1000$ \ce{W/cm^2} with $E_\lambda$ being the solar spectral irradiance of the AM1.5 global solar spectrum. The heat transfer coefficient $h$ at the water-to-air interface in the fabric is about 60 $W/m^2/K$ \cite{12}.

As described in Ref. \cite{6}, for the floating photothermal fabric, the optical energy is mainly harvested by a water layer of 3 cm below the PANi-cotton membrane. The membrane plays a role of antireflection coating and the absorbed sunlight on the fabric is supposed to quickly pass into water. Ref. \cite{6} reports that only 20 $\%$ of sunlight is passed through the system \cite{6}. Thus, $P_0\approx \int_{300}^{2500}E_\lambda(1-e^{-\alpha_w(\lambda) L_2}) d\lambda \approx 256.4$ \ce{W/cm^2} with $\alpha_w(\lambda)$ being the absorption coefficient of the water solution \cite{30}. The absorbed energy quantitatively agrees with Ref. \cite{6}. The heat transfer coefficient $h$ between liquid and a substrate is $\sim$ 1000 $W/m^2/K$ \cite{12}.

%\textcolor{red}{Note that}
%%%%%%%%%%%%%%%%%%%%%%%%%%%%%%%%%%%%%%%%%%%%%%
%%%%%%%%%%%%%%%%%%%%%%%%%%%%%%%%%%%%%%%%%%%%%%
\section{Results and discussions}
\begin{figure}[htp]
\includegraphics[width=9cm]{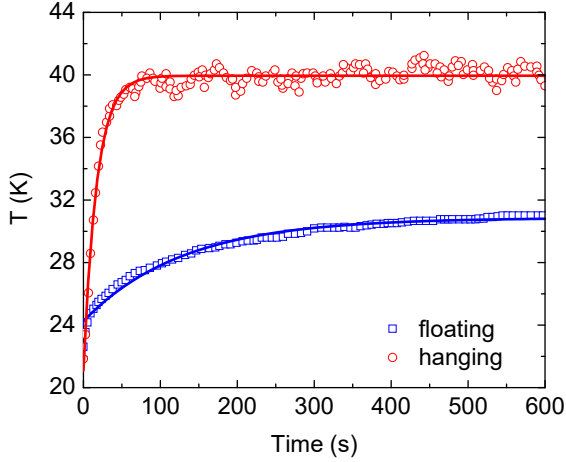}
\caption{\label{fig:2}(Color online) Experimental \cite{6} and theoretical (Eq. (\ref{eq:10})) temperature at the surface ($z=0$) of the hanging and floating PANi-cotton membrane as a function of time under sunlight irradiation. Discrete data points and solid curves correspond to the experimental data and theoretical calculations, respectively.}
\end{figure}

We first consider how the temperature in both PANi-cotton fabrics changes in time. For this purpose we utilize an exponentially decaying function found from solving an energy balance equation in homogeneous and open systems under light irradiation \cite{2,3}. This analytical expression is typically applied to a certain wavelength light, however, it is expected that such a solution is applicable in the cases studied here where light is a superposition of wavelengths. The time dependent temperature function is 
\begin{eqnarray}
T(t)=T_{max}-(T_{max}-T_0)e^{-Bt}
\label{eq:10}
\end{eqnarray}
where $T_{max}$ and $T_0$ are the maximum and medium temperatures in the photothermal process, respectively. Also, the exponential decay constant is independent of the incident sunlight wavelengths and it given as $B=hS/m_sC_s$, where  $S$ is a surface area of the container, $m_s$ and $C_s$ are the mass and heat capacity of a solution, respectively, as reported in \cite{2,3}. The wavelength-dependent absorption affects only $T_{max}$. After fitting Eq. (\ref{eq:10}) with the experimental data in \cite{6} for the average surface temperatures of the hanging and floating PANi-cotton fabric under solar illumination, we find: (1) for the hanging photothermal fabric, $T_{max}=39.95$ $^0C$, $T_{0}=21.1$ $^0$C, and $B = 0.0578$; (2) for the floating photothermal fabric, $T_{max}=30.86$ $^0$C, $T_{0}=24.25$ $^0$C, and $B =0.00776$. The experimental data points and the theoretical curves are shown in Fig. \ref{fig:2}. The medium temperature ($T_{0}=24.25$ $^0$C) in the case of the floating membrane is higher when compared to the measured one ($\approx 21.5$ $^0$C). A reason may be due to the measurement process in experiment. The experimental data gives $T(t=0)=22.7$ $^0$C and this leads to sequential deviation. However, overall, the theoretical fittings show a very good agreement with the experimental results.

\begin{figure}[htp]
\includegraphics[width=8.5cm]{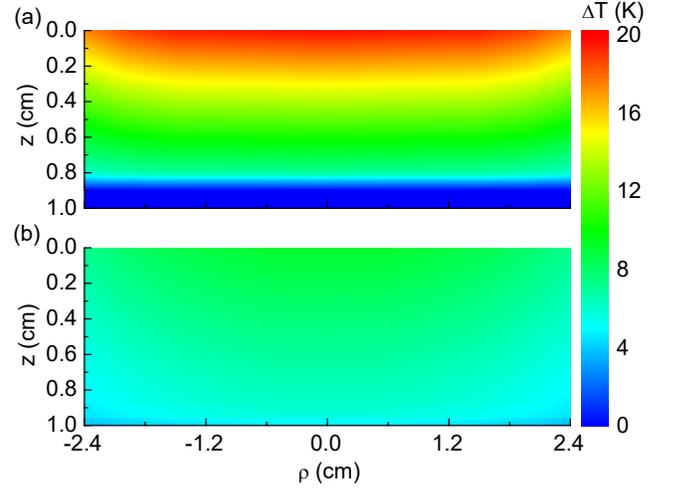}
\caption{\label{fig:3}(Color online) Spatial contour plots of the steady-state temperature increase in Kelvin units in (a) hanging and (b) floating PANi-cotton fabrics calculated by Eqs. (\ref{eq:8}), (\ref{eq:11}), and (\ref{eq:9}) with parameters given in the caption of Fig. \ref{fig:1}.}
\end{figure}

The spatial distribution of the steady-state temperature is also calculated using Eqs. (\ref{eq:8}), (\ref{eq:11}), and (\ref{eq:9}) for the hanging and floating PANi-cotton fabric in the absence of nanoparticles ($N = 0$) when exposed under solar irradiation. Numerical results are shown in Fig. \ref{fig:3} using parameters from the experimental report in \cite{6}. We find that $\Delta T(z)$ significantly changes in the hanging fabric and the photothermal conversion process is highly localized within the membrane because of the rapid decay in the outside air region. We also obtain that the average temperature at the surface of the hanging PANi-cotton fabric is about 41 $^0$C (averaging $T(\rho,z=0)$ with respect to $\rho$ near the center of surface). In the case of the floating PANi-cotton membrane, the calculated surface temperature is approximately 30.7 $^0$C, which is very close to $\approx$ 30.86 $^0$C as shown in Fig. \ref{fig:2} and Ref. \cite{6}. Since the vaporized weight of water is essentially determined by the temperature discrepancy at the liquid-air surface \cite{31}, the water in the hanging fabric vaporizes faster than that in the floating counterpart. These are quantitatively consistent with prior experiments \cite{6}.

In the presence of TiN nanoparticles, the absorbed light energy and the absorption coefficient of the solution are increased. We consider the case of dilute aqueous nanoparticle solution whose absorption coefficient is given by the Beer-Lambert law \cite{31}
\begin{eqnarray}
\alpha(\omega) =\alpha_w(\omega) + N Q_{ext},
\label{eq:16}
\end{eqnarray}
where $Q_{ext}$ is the extinction cross section theoretically calculated by Mie scattering theory \cite{10}, which is
\begin{eqnarray}
Q_{ext} &=& \frac{2\pi}{x^2}\sum_{n=1}^{\infty}(2n+1){Re}\left(a_n + b_n \right), 
\label{eq:ext}
\end{eqnarray}
where
\begin{eqnarray}
a_n &=& \frac{m^2j_n(mx)\left[xj_n(x)\right]' - \mu j_n(x)\left[mxj_n(mx)\right]'}{m^2j_n(mx)\left[xh^{(1)}_n(x)\right]' - \mu h^{(1)}_n(x)\left[mxj_n(mx)\right]'} \nonumber\\
b_n &=& \frac{\mu j_n(mx)\left[xj_n(x)\right]' -j_n(x)\left[mxj_n(mx)\right]'}{\mu j_n(mx)\left[xh^{(1)}_n(x)\right]' -h^{(1)}_n(x)\left[mxj_n(mx)\right]'}
\label{eq:ext-1}
\end{eqnarray}
where $x = kR$, $k=\omega\sqrt{\varepsilon_w(\omega)}/c$ is the wavenumber in water, $c$ is the speed of light, $\varepsilon_w(\omega)$ is a dielectric function of water, $j_n(x)$ is the spherical Bessel function of the first kind, $h_n^{(1)}(x)$ is the spherical Hankel function of the first kind, $m = \sqrt{\varepsilon(\omega)/\varepsilon_w(\omega)}$, and $\varepsilon(\omega)$ is the dielectric function of TiN.

The complex dielectric function of TiN is fitted by a generalized Drude-Lorentz model \cite{31,11} with experimental data for thin film. The analytical expression of this model is
\begin{eqnarray}
\varepsilon(\omega)=\varepsilon_{\infty}-\frac{\omega_{p}^2}{\omega(\omega+i\Gamma_D)}+\sum_{j=1}^2\frac{\omega_{L,j}^2}{\omega_{0,j}^2-\omega^2-i\gamma_j\omega},
\label{eq:3}
\end{eqnarray}
where $\varepsilon_{\infty}=5.18$ is the high-frequency permittivity, $\omega_{p} \approx 7.38$ eV is the plasma frequency, and $\Gamma_D \approx 0.26$ eV is the Drude damping parameter \cite{11}. Also, $\omega_{L,1}\approx 6.5$ \ce{eV} and $\omega_{L,2}=1.5033$ \ce{eV} are the Lorentz oscillator strengths, $\omega_{0,1}=4.07$ \ce{eV} and $\omega_{0,2}=2.02$ \ce{eV} are the Lorentz energies, and $\gamma_1 = $ 1.42 \ce{eV} and $\gamma_2 = $ 0.87 \ce{eV} are the Lorentz damping parameters \cite{11}. After calculating $Q_{ext}$ by Eqs. (\ref{eq:ext}) and (\ref{eq:ext-1}), the absorbed energy is found as
\begin{eqnarray}
P_0\approx \int_{300}^{2500}E_\lambda\left(1-e^{-[\alpha_w(\lambda) + N Q_{ext}]L_2}\right) d\lambda.
\label{eq:4}
\end{eqnarray}

\begin{figure}[htp]
\includegraphics[width=9cm]{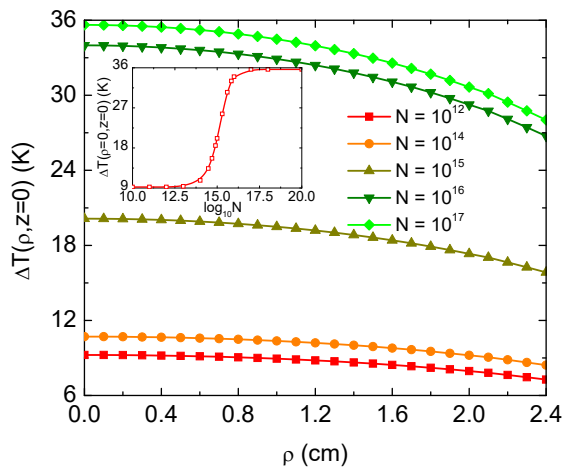}
\caption{\label{fig:4}(Color online) Temperature distribution at the surface (calculated by Eqs. (\ref{eq:8}), (\ref{eq:11}), (\ref{eq:9}), and (\ref{eq:4}) with parameters given in the caption of Fig. \ref{fig:1}) of the floating PANi-cotton fabric at different densities of TiN nanparticles of $R = 50$ nm in unit of \ce{nanoparticles/m^3}. The inset shows the maximum surface temperature as a function of the density number.}
\end{figure}

The spatial distribution of the surface temperature for the floating PANi-membrane above an aqueous TiN nanoparticle solution is calculated using Eqs. (\ref{eq:8}), (\ref{eq:11}), and (\ref{eq:9}) with parameters for the absorbed energy and Eq. (\ref{eq:4}) as described above. Results for $\Delta T(\rho,z=0)$ for different concentrations of nanoparticles with 50 nm radius are shown in Fig. \ref{fig:4}. At the hanging edges less incident light hits the water surface, thus there is greater decrease in $\Delta T(\rho,z=0)$ when compared with the center of the membrane. The temperature at the hottest spot  of the system ($\rho=z=0$) as a function of $\log_{10}N$ is shown in the inset. Given that the TiN nanofluid is a dilute nanoparticle solution, the optical energy absorbed by the nanoparticles themselves plays a minor role in the effective absorption process. For the case of $N\leq 10^{13}$ \ce{particles/m^3}, the temperature gradient is almost constant such that $\Delta T(\rho,z=0) \approx 9.2$ $K$.

As the concentration is increased, more electromagnetic energy is trapped inside the solution due to increased optical absorption. This leads to an overall increased of the temperature gradient. Particularly, $\Delta T(\rho,z=0) \geq 20$ $K$ as $N \geq 10^{15}$. When $N \geq 10^{18}$, this average surface temperature can increase by 36 $K$, which is about 1.8 times larger than for the hanging fabric whose $\Delta T (\rho,z=0) \approx 19.6$ $K$. At a given nanoparticle density, the surface temperature can be increased by extending the effective projected size of fabrics $w$. Based on Eq. (\ref{eq:9}), we find that the heat flux $\Psi(u)$ grows with $w$ and it raises $\Delta T (\rho,z=0)$ \cite{4}.

Considering the weight of vaporized water $\Delta m$, which is linearly proportional to $\Delta T(\rho,z=0)$ \cite{31}, we determine that the evaporation process using the hanging fabric can be nearly surpassed by the floating membrane with the help of dispersed plasmonic nanoparticles with an appropriate density. For $N \geq 10^{18}$, the solution has sufficient amount of nanoparticles to absorb total sunlight. In this case, the heat source remains nearly unchanged and the mass of water is much larger than that of nanoparticles. Thus, $\Delta T(\rho=0,z=0)$ becomes saturated as seen in the inset of Fig. \ref{fig:4}.

\begin{figure}[htp]
\includegraphics[width=9cm]{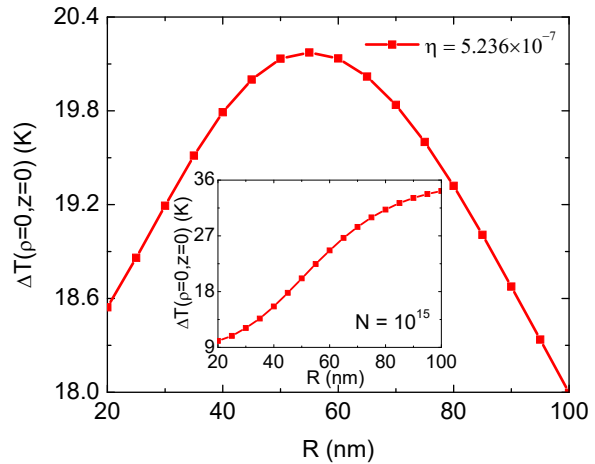}
\caption{\label{fig:5}(Color online) The maximum temperature of $\Delta T$ of the floating PANi-cotton fabric as a function of nanoparticle radius when the volume fraction $\eta = 5.236\times10^{-7}$ is fixed. 
The inset shows the same functional dependence at a fixed nanoparticle density $N=10^{15}$ particles/\ce{m^3} (inset).}
\end{figure}

The photothermal process for the floating membrane can further be tuned by changing the size of the TiN nanoparticles. In Fig. \ref{fig:5} we show results for $\Delta T(\rho=z=0)$ of the floating PANi-cotton fabric calculated using Eqs. (\ref{eq:8}), (\ref{eq:11}), (\ref{eq:9}), and (\ref{eq:4}) by keeping the concentration and volume fraction constants. It is noted in this case that despite the presence of many particles with small $R$, their individual optical extinction is small. At the same time, larger particles have greater extinction cross section but the number of particles is small. This trade-off between number and size of nanoparticles affects $T(\rho=z=0)$ significantly, such that the collective heating process exhibits its best performance when $R$ ranges from 50 nm to 60 nm, as seen in Fig. \ref{fig:5}. It is found that this is characteristic behaviour for larger concentrations. However, for  $N \leq 10^{15}$ increasing the particle size leads to an increase of $Q_{ext}$. This type of monotonic growth of $\Delta T(\rho=z=0)$ with the nanoparticle radius is shown in the inset of Fig. \ref{fig:5}. Prior work \cite{29} suggests that $Q_{ext}$ of a single particle is altered by its neighbors as the nearest neighbor interparticle separation is less than $4R$. Thus, for $N \geq 1/(4R)^3$ or $\eta \geq 0.0654$, $Q_{ext}$ changes.

\section{Conclusion}
We have investigated photothermal process in hanging and floating above water PANi-cotton fabrics under solar illumination. For this purpose, we consider the heat transfer differential equation to numerically calculate the temperature gradient in various cases. The model is validated by available experimental data in \cite{6}. 
The PANi-cotton fabric is found to be an anti-reflective coating in the frequency range from 300 nm to 2500 nm \cite{6}. Thus, the hanging fabric absorbs the total light flux and perfectly converts it into heat which results in temperature increase. The floating fabric also absorbs 100 $\%$ of the solar energy but the absorbed energy is quickly transferred into the underneath water layer. Thus, the solar thermal heating of the floating system is less effective than that of the hanging one.

To enhance the photothermal effects of the floating fabric, we consider random dispersion of TiN nanoparticles in water. Increasing the nanoparticle density significantly enhances the energy absorption in solutions and reduces the penetration depth of sunlight, which further leads to an increased temperature profile. Since the vaporized weight of water is proportional to the temperature deviation at a liquid-air interface \cite{31},  photothermal effects and solar evaporation in the case of the floating fabric can be much improved surpassing the performance of  the hanging fabric. The density and size of TiN nanoparticles can also be used for further modulations of the surface temperature. The approach gives an excellent tool for designing effective applications related to solar photothermal heating.

\begin{acknowledgments}
This research is funded by Vietnam National Foundation for Science and Technology Development (NAFOSTED) under grant number 103.01-2018.337. This work was supported by JSPS KAKENHI Grant Numbers JP19F18322 and JP18H01154. L.M.W. acknowledges financial support from the US Department of Energy under Grant No. DE-FG02-06ER46297. 

Conflict of Interest: The authors declare that they have no conflict of interest.
\end{acknowledgments}


\begin{thebibliography}{5}
\bibitem{17} K. Bae, G. Kang, S. K. Cho, W. Park, K. Kim, and W. J. Padilla, Flexible thin-film black gold membranes with ultrabroadband plasmonic nanofocusing for efficient solar vapour generation. \emph{Nat. Commun.} {\bf 2015}, 6, 1-9.
\bibitem{18} T. S. Kao, S. D. Jenkins, J. Ruostekoski, and N. I. Zheludev, Coherent Control of Nanoscale Light Localization in Metamaterial: Creating and Positioning Isolated Subwavelength Energy Hot Spots. \emph{Phys. Rev. Lett.} {\bf 2011}, 106, 085501.
\bibitem{21} S. Lal, S. E. Clare and N. J. Halas, Nanoshell-enabled photothermal cancer therapy: impending clinical impact. \emph{Acc. Chem. Res.} {\bf 2008}, 41, 1842-1851.
\bibitem{22} V. T. T. Duong, A. D. Phan, N. T. H. Lien, D. T. Hue, D. Q. Hoa, D. T. Nga, T. H. Nhung and N. A. Viet, Near-Infrared Photothermal Response of Plasmonic
Gold-Coated Nanoparticles in Tissues. \emph{Phys. Status Solidi A} {\bf 2018},  215, 1700564.
\bibitem{23} J.-W. Xu, K. Yao, and Z.-K. Xu, Nanomaterials with a photothermal effect for antibacterial activities: an overview. \emph{Nanoscale} {\bf 2019}, 11, 8680-8691.
\bibitem{19} E. Rephaeli, A. Raman, S. Fan, Ultrabroadband Photonic Structures To Achieve High-Performance Daytime Radiative Cooling. \emph{Nano Lett.} {\bf 2013}, 13, 1457-1461.
\bibitem{24} X. Huang, I. H. El-Sayed, W. Qian, M. A. El-Sayed, Cancer Cell Imaging and Photothermal Therapy in the Near-Infrared Region by Using Gold Nanorods. \emph{J. Am. Chem. Soc.} {\bf 2006}, 128, 2115-2120.
\bibitem{25} R. Huschka, J. Zuloaga, M. W. Knight, L. V. Brown, P. Nordlander, N. J. Halas, Light-Induced Release of DNA from Gold Nanoparticles: Nanoshells and Nanorods. \emph{J. Am. Chem. Soc.} {\bf 2011}, 133, 12247-12255.
\bibitem{6} Z. Liu, B. Wu, B. Zhu, Z. Chen, M. Zhu, and X. Liu, Continuously Producing Watersteam and Concentrated Brine from Seawater by Hanging Photothermal Fabrics under Sunlight. \emph{Adv. Funct. Mater.} {\bf 2019}, 29, 1905485.

\bibitem{31} A. D. Phan, N. B. Le, T. H. L. Nghiem, L. M. Woods, S. Ishii, and K. Wakabayashi, Confinement effects on the solar thermal heating process of TiN nanoparticle solutions. \emph{Phys. Chem. Chem. Phys.} {\bf 2019}, 21, 19915.
\bibitem{26} C. Ma, J. Yan, Y. Huang, C. Wang, and G. Yang, The optical duality of tellurium nanoparticles for broadband solar energy harvesting and efficient photothermal conversion. \emph{Sci. Adv.} {\bf 2018}, 4, eaas9894.
\bibitem{16} A. Guo, Y. Fu, G. Wang, and X. Wang, Diameter effect of gold nanoparticles on photothermal conversion for solar steam generation. \emph{RSC Adv.} {\bf 2017}, 7, 4815-4824.
\bibitem{7}  F. S. Awad, H. D. Kiriarachchi, K. M. AbouZeid, U. Ozgur, and M. S. El-Shall, Plasmonic Graphene Polyurethane Nanocomposites for Efficient Solar Water Desalination. \emph{ACS Appl. Energy Mater.} {\bf 2018}, 1, 976-985.
\bibitem{15} A. D. Phan, N. B. Le, N. T. H. Lien, and K. Wakabayashi, Multilayered Plasmonic Nanostructures for Solar Energy Harvesting. J. Phys. Chem. C {\bf 2018}, 122, 19801-19806.
\bibitem{32} X. Wang, Y. He, X. Liu, G. Cheng and J. Zhu, Appl. Energy, {\bf 2017}, 195, 414-425.
\bibitem{33} K.-K. Liu, Q. Jiang, S. Tadepalli, R. Raliya, P. Biswas, R. R. Naik, and S. Singamaneni, ACS Appl. Mater. Interfaces {\bf 2017}, 9, 7675-7681.
\bibitem{28} O. Neumann, A. S. Urban, J. Day, S. Lal, P. Nordlander, and N. J. Halas, Solar vapor generation enabled by nanoparticles. ACS Nano {\bf 2013}, 7, 42-49.
\bibitem{1} A. Sousa-Castillo, O. Ameneiro-Prieto, M. Comesana-Hermoa, R. Yu, J. M. Vila-Fungueirino, M. Perez-Lorenzo, F. Rivadulla, F. J. G. de Abajo, M. A. Correa-Duarte, Hybrid plasmonic nanoresonators as efficient solar heat shields. Nano Energy {\bf 2017}, 37, 118-125.
\bibitem{34} S. Zhang, X. Xu, T. Lin, and P. He, Recent advances in nano-materials for packaging of electronic devices, J. Mater. Sci.: Mater. Electron {\bf 2019}, 30, 13855-13868.
\bibitem{35} T. Jin, Y. Pan, G.-J. Jeon, H.-I. Yeom, S. Zhang, K.-W. Paik, and S.-H. Ko Park, Ultrathin Nanofibrous Membranes Containing Insulating Microbeads for Highly Sensitive Flexible Pressure Sensors, ACS Appl. Mater. Interfaces {\bf 2020}, 12, 13348-13359.
\bibitem{36} A disulfiram-loaded electrospun poly(vinylidene fluoride) nanofibrous scaffold for cancer treatment, Nanotechnology {\bf 2020}, 31, 115101.
\bibitem{9} \url{https://www.memsnet.org/material/titaniumnitridetinbulk/}
\bibitem{11} H. Reddy, U. Guler, Z. Kudyshev, A. V. Kildishev, V. M. Shalaev, and A. Boltasseva, Temperature-Dependent Optical Properties of Plasmonic Titanium Nitride Thin Films. ACS Photonics {\bf 2017}, 4, 1413–1420.
\bibitem{27} G. V. Naik, J. Kim, and A. Boltasseva, Oxides and nitrides as alternative plasmonic materials in the optical range. Opt. Mater. Express {\bf 2011}, 1, 1090-1099.
\bibitem{8} A. D. Phan, D. T. Nga, D. C. Nghia, V. D. Lam, and K. Wakabayashi, Effects of Mid‐Infrared Graphene Plasmons on Photothermal Heating. \emph{Phys. Status Solidi RRL} {\bf 2020}, 14, 1900656.
\bibitem{12} Y. A. Cengel, \emph{Heat Transfer: A Practical Approach} (McGrawHill: Boston, 1998).
\bibitem{30} G. M. Hale and M. R. Querry, Optical Constants of Water in the 200-$nm$ to 200-$\mu m$ Wavelength Region. \emph{Appl. Opt.} {\bf 1973}, 12, 555-563.
\bibitem{2} C. M. Hessel, V. P. Pattani, M. Rasch, M. G. Panthani, B. Koo, and J. W. Tunnell, Copper Selenide Nanocrystals for Photothermal Therapy. Nano Lett. {\bf 2011}, 11, 2560-2566.
\bibitem{3} D. K. Roper, W. Ahn, and M. Hoepfner, Microscale Heat Transfer Transduced by Surface Plasmon Resonant Gold Nanoparticles. J. Phys. Chem. C {\bf 2007}, 111, 3636-3641.
\bibitem{10} M. Quinten, \emph{Optical Properties of Nanoparticle Systems}, (Wiley, Weinheim, Germany, 2011).
\bibitem{4} See Supplemental material for information regarding how $\Delta T(\rho=z=0)$ increases with the size of the photothermal floating fabrics.
\bibitem{29} G. Domingues, S. Volz, K. Joulain, and J.-J. Greffet, Phys. Rev. Lett. {\bf 2005}, 94, 085901.






\end{thebibliography}
\end{document}